\documentclass{osa-article}

\journal{osajournal}
\articletype{Research Article}
\usepackage{tikz,xcolor,hyperref}

\definecolor{lime}{HTML}{A6CE39}

\begin{document}

\title{A packaged whispering gallery resonator device based on an optical nanoantenna coupler}

\author{Angzhen Li,\authormark{1,2} Ke Tian,\authormark{2} Jibo Yu,\authormark{1,2} Rashmi A. Minz,\authormark{3} Jonathan M. Ward,\authormark{4} Samir Mondal,\authormark{3} Pengfei Wang,\authormark{1,5,*} and S\'{i}le Nic Chormaic\authormark{2,+}}

\address{\authormark{1}Key Laboratory of In-Fiber Integrated Optics of Ministry of Education, College of Science, Harbin Engineering University, Harbin 150001, China\\
\authormark{2}Light-Matter Interactions for Quantum Technologies Unit, Okinawa Institute of Science and Technology Graduate University, Onna, Okinawa 904-0495, Japan\\
\authormark{3}Central Scientific Instruments Organisation (Council of Scientific and Industrial Research, India), Chandigarh 160030, India\\
\authormark{4}Physics Department, University College Cork, Cork, Ireland\\
\authormark{5}Key Laboratory of Optoelectronic Devices and Systems of Ministry of Education and Guangdong Province College of Optoelectronic Engineering, Shenzhen University, Shenzhen 518060, China\\
}

\email{\authormark{*}pengfei.wang@tudublin.ie}
\email{\authormark{+}sile.nicchormaic@oist.jp}
%% email address is required



\begin{abstract}
In this work, we present the design and fabrication of a packaged whispering gallery mode (WGM) device based on an optical nanoantenna as the coupler and a glass microsphere as the resonator. The microspheres were fabricated from SiO$_2$ fiber or Er$^{3+}$-doped fiber, the latter creating a WGM laser with a threshold of 93 $\mu$W at 1531 nm. The coupler-resonator WGM device is packaged in a glass capillary. The performance of the packaged microlaser is characterized, with lasing emission  both excited in and collected from the WGM cavity via the nanoantenna. The packaged system provides isolation from environmental contamination, a small size, and unidirectional coupling while maintaining a high quality (Q-) factor ($\sim$10$^8$). It opens up new possibilities for practical applications of WGM microdevices in a variety of fields such as low threshold lasers, filters, and sensors.
\end{abstract}

%%%%%%%%%%%%%%%%%%%%%%%%%%  body  %%%%%%%%%%%%%%%%%%%%%%%%%%
\section{Introduction}

 Although whispering gallery mode (WGM) devices have shown great potential in many fields \cite{PhysRevLett.101.093902,2009On, PhysRevLett.106.113901,monifi2013tunable,Yang:16, Fang:17,Ward:18, doi:10.1021/acsnano.9b04702}, they have encountered obstacles in practical applications. The optical coupler, a tool for transferring the light in and out of the resonators, is a very important part of the WGM system. Among the various coupling schemes, such as prisms \cite{doi:10.1063/1.1540242} and angle-polished fibers \cite{10.1117/12.2519229}, tapered fibers \cite{Knight:97} have been demonstrated to be ideal because of their high coupling efficiency and ability to achieve critical coupling \cite{PhysRevLett.85.74}. They can also be used to tailor the cavity input-output relations in the WGM resonator-waveguide system \cite{PhysRevLett.124.103902}. However, there are still issues with the tapered fiber coupling scheme, especially in terms of practical applications. First, tapered fibers are typically less than a few $\mu$m in diameter, making them extremely fragile. Second, the resonator-coupler system can be affected by environmental factors such as air currents, which continuously and irregularly change the coupling state. Dust in the environment can also significantly reduce the quality (Q)-factors and efficiency of the coupling. A packaging scheme using two glass plates and a tube to fix the microsphere and tapered fiber can make the microsphere-tapered fiber system move as a whole while maintaining the Q-factor (\begin{math}1.08\times10^8\end{math} at 1550 nm). However, it does not solve the problems of vulnerability and environmental pollution \cite{Dong:15}. Packaging resonator-tapered fiber systems with low refractive index optical glues have been shown to go some way toward solving the problem of damage and contamination \cite{Yan:11}. A low-index polymer packaged microtoroid-tapered fiber device with a Q-factor up to \begin{math}2\times10^7\end{math} at 780 nm has been demonstrated \cite{kavungal2017a}. However, reports have shown that this method can degrade the Q-factors of the system \cite{Yan:11,7515011,kavungal2017a,ZHAO2017875}. Also, limited by the length of the tapered fiber, the size of such resonator-tapered fiber packaged devices is often on the centimeter scale \cite{Dong:15,kavungal2017a,ZHAO2017875}. Another idea  to further scale down the devices is to place the resonator in a capillary tube to form an embedded WGM device. For example, excitation and collection of the whispering gallery modes can be achieved using an etched thin-walled capillary \cite{Bai:18}, a tapered hollow annular core fiber \cite{Wang:18}, embedded dual core hollow fiber \cite{Zhang:18} or a femtosecond laser-engraved waveguide \cite{Liu:20} as a coupler. These embedded packaging solutions offer advantages in terms of system stability, smaller size, and immunity to environmental interference, but the Q-factors can be seriously affected.   

 Our previous work \cite{Ward:19} demonstrated coupling to WGM resonators using a nanoantenna based on Rayleigh scattering \cite{shu2018scatterer}. The nanoantenna effectively excites the cavity to form WGM resonances and maintain high Q-factors. Good coupling efficiency compared with other coupling methods based on Rayleigh scattering can also be achieved. In this paper, we propose and demonstrate a packaged WGM resonator with a fiber-based nanoantenna. This small footprint packaged device addresses environmental contamination while maintaining very high Q-factors of the order of $10^8$ at 1531~nm pump. In addition, the nanoantenna structure allows for excitation and collection of signals using a single fiber probe, showing considerable promise in terms of practicality. Based on this scheme, we have also fabricated and packaged an Er$^{3+}$-doped WGM laser and the related performances are studied and discussed. The demonstration of lasing in a packaged WGM device fabricated directly from Er-doped fiber is important and avoids the necessity of coating via solgel \cite{Yang:03, Yang:17} or using other methods \cite{4447284, CHEN20093765, Ward2016sr}.

\section{Device Fabrication}

\begin{figure}[htbp]
\centering\includegraphics[width=8cm]{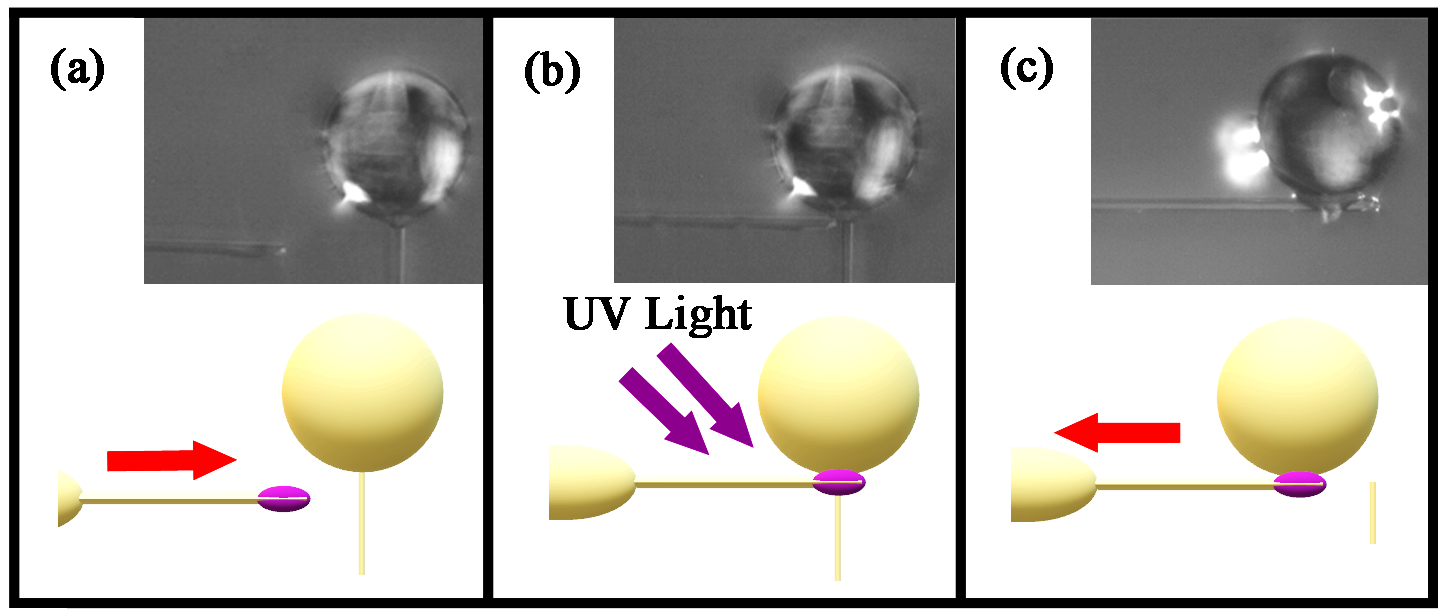}
\caption{Mounting a microsphere to a connecting rod. (a) Approach a half-tapered optical fiber coated with UV glue to the connection point between the microsphere and the connecting rod. (b) Fix the half-taper at the connection point under UV light. (c) Tear the microsphere free from the original fiber rod by adjusting the position of the 3D stage.}
\label{fig:fabrication of sphere}
\end{figure}

There are several techniques used to make spherical WGM resonators \cite{10.1117/12.323384,CHEN20093765, PhysRevLett.85.74, doi:10.1063/1.3455198,Madugani:12}. One of the more common methods involves melting the end of a section of optical fiber, using, for example, a CO$_2$ laser \cite{PhysRevLett.85.74} and we used this technique for the initial microsphere fabrication. Such microspheres have a fiber rod at the pole perpendicular to the equatorial plane where the resonance is formed. Although this rod does not affect the formation of the resonance at the equator, it poses difficulties for packaging and miniaturization of the device. To overcome this, here, we instigated an additional fabrication step in order to have a half-tapered fiber as a manipulation rod parallel to the resonance-forming surface. Figure 1 illustrates the process. The half-tapered fiber, with a small amount of UV curing glue on the tip, and a microsphere were fixed on two 3D stages. Under a microscope, the two 3D stages were adjusted to bring the tip of the half-tapered fiber into contact with the connecting point of the microsphere and its thin fiber rod (Fig. \ref{fig:fabrication of sphere}(a)). Next, the contact area was irradiated with UV light for a few seconds to cure the UV glue (Fig. \ref{fig:fabrication of sphere}(b)). After curing, the microsphere was torn from its original rod by moving the 3D stage. For this to work, the half-tapered fiber must be thicker than the microsphere's intrinsic fiber rod to facilitate breakage when moving the stage. Now the microsphere has a supporting rod parallel to the resonance plane (Fig. \ref{fig:fabrication of sphere}(c)). In this experiment, the fiber used to prepare both the microsphere and the half-tapered fiber was a commercial silica fiber (1060XP, Thorlabs).

\begin{figure}[htbp]
\centering\includegraphics[width=6cm]{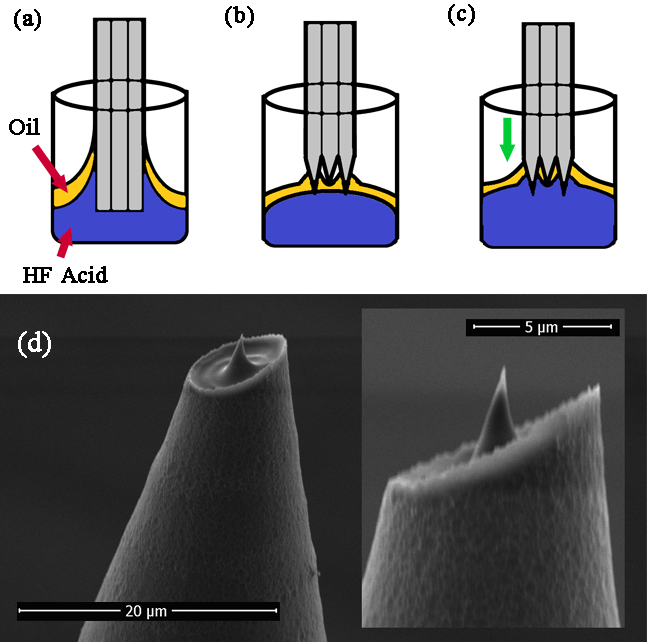}
\caption{Fabrication of a nanoantenna. (a) The end face of the fiber is brought into contact with the HF acid surface. Under the action of liquid surface tension, HF acid rises up along the outer wall of the fiber and etches the cladding and core. (b) After 30 minutes, the core inside the fiber forms a tiny tip. Because the core is etched faster than the cladding, the sample has a ring tube formed by cladding that is longer than the tip. (c) Moving the sample down 20 $\mu$m and continue etching for 10 minutes. (d) A scanning electron microscope (SEM) image of a fabricated nanoantenna.}
\label{fig:probe}
\end{figure}

To fabricate the nanoantenna for optical coupling to the microsphere, we chemically etched photosensitive optical fiber (GF3, Nufern), using a technique developed by Mondal \textit{et al.}, as reported in \cite{Mondal:09,5934361}. For this particular fiber, the fiber core etches quicker than the cladding under the same hydrofluoric (HF) acid etching conditions. The first step in the preparation process was to remove the coating from one end of the fiber and cut the end face flat. It was then secured in a vertical position with a micrometer-adjustable fiber optic holder. A small plastic vial containing 48$\%$ HF acid solution was fixed directly below the optical fiber. An oil film on the surface of the acid solution prevented evaporation. The etching process was divided into two steps. First, the position of the fiber holder was adjusted so that the end face of the fiber just touched the surface of the acid. Under the action of liquid surface tension, the acid rose up along the outer wall of the fiber and etched the cladding and core simultaneously (Fig. \ref{fig:probe}(a)). After 30 minutes, the core inside the fiber formed a tiny tip (Fig. \ref{fig:probe}(b)). The second step involved adjusting the holder to move the fiber down into the acid by about 20 $\mu$m (Fig. \ref{fig:probe}(c)). The elevated height of the acid level and the difference between the etching rates of the cladding and core materials created a capillary ring at the core-cladding boundary.

\begin{figure}[htbp]
\centering\includegraphics[width=7cm]{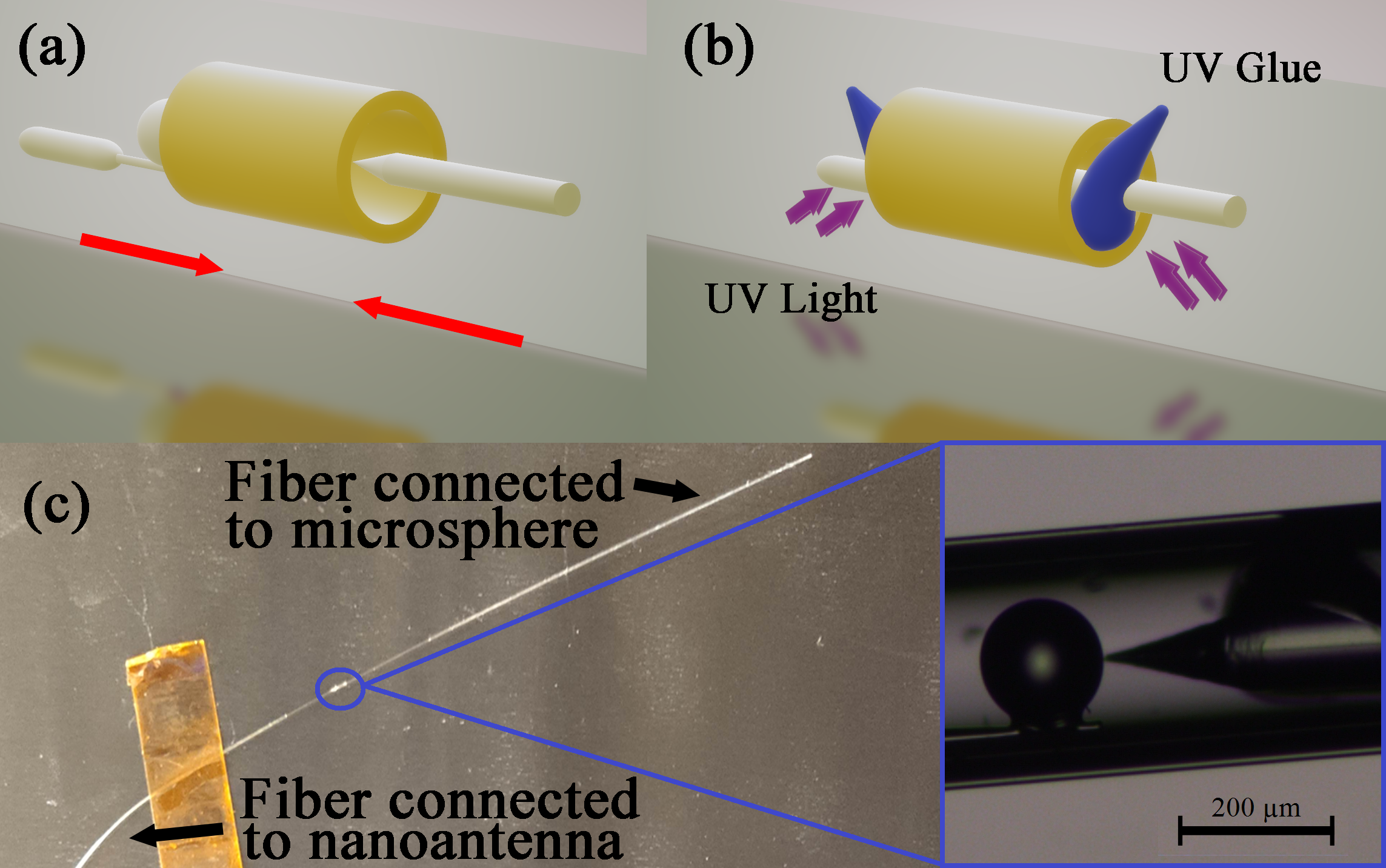}
\caption{Packaging process. (a) The microsphere and nanoantenna are inserted into a short capillary tube and adjusted in position. (b) The two sides of the capillary are sealed with UV glue and irradiated by UV light (c) A packaged device. The zoomed-in image shows a packaged structure in detail.}
\label{fig:packaging}
\end{figure}

After about 10 minutes of etching, an antenna with a protruding nanoprobe was obtained. The etched sample was cleaned with acetone to remove the oil layer attached to the surface of the nanoantenna. A scanning electron microscope (SEM) image of a fabricated nanoantenna is shown in Fig. \ref{fig:probe}(d). Finally, the nanoantenna and the microsphere were inserted into a capillary using 3D stages (Fig. \ref{fig:packaging}(a)). After adjusting their relative positions under a microscope, while monitoring the WGM spectrum, the two sides of the capillary were sealed with UV glue and irradiated by UV light (Fig. \ref{fig:packaging}(b)). Figure \ref{fig:packaging}(c) shows a prepared sample with a microsphere diameter of 160 $\mu$m packaged in a capillary with 350 $\mu$m outer diameter and 250 $\mu$m inner diameter. The fiber on one side of the nanoantenna acts as a waveguide for the output and input optical signals, while the pigtail fiber on the other side, which is connected to the microsphere, can be cut off as needed.

\section{Results and Discussion}

\begin{figure}[htbp]
\centering\includegraphics[width=8cm]{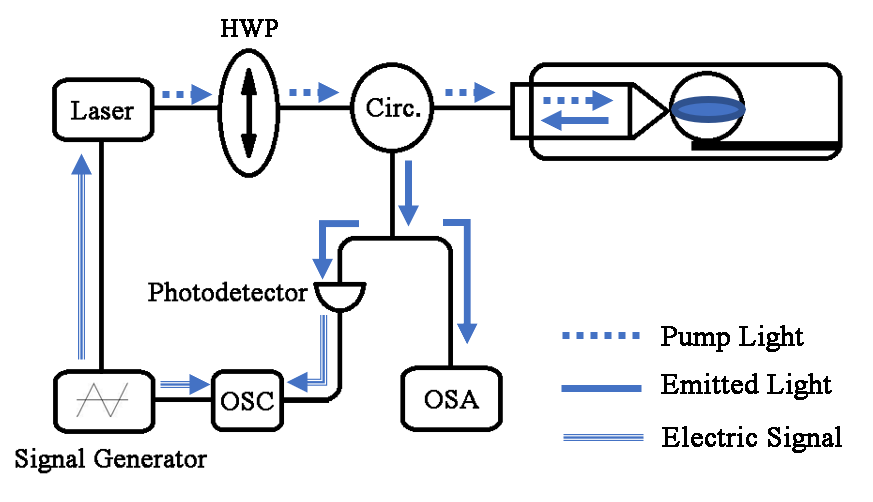}
\caption{Schematic of the test system. OSC: Oscilloscope; OSA: Optical spectrum analyzer; HWP: Halfwave plate; Circ.: Circulator.}
\label{fig:setup}
\end{figure}

A schematic of the experimental setup to test the performance of the packaged devices is shown in Fig. \ref{fig:setup}. Pump light at 1531~nm was coupled to the microsphere via the nanoantenna. A 3-port circulator was used to separate the input and collected signals. A half-wave plate on the input beam was used for polarization adjustment. The signal was transmitted from the output of the circulator and observed on an oscilloscope (WaveSurfer 10, Teledyne LeCroy) and an optical spectrum analyzer (MS9740B, Anritsu).

\begin{figure}[htbp]
\centering\includegraphics[width=11cm]{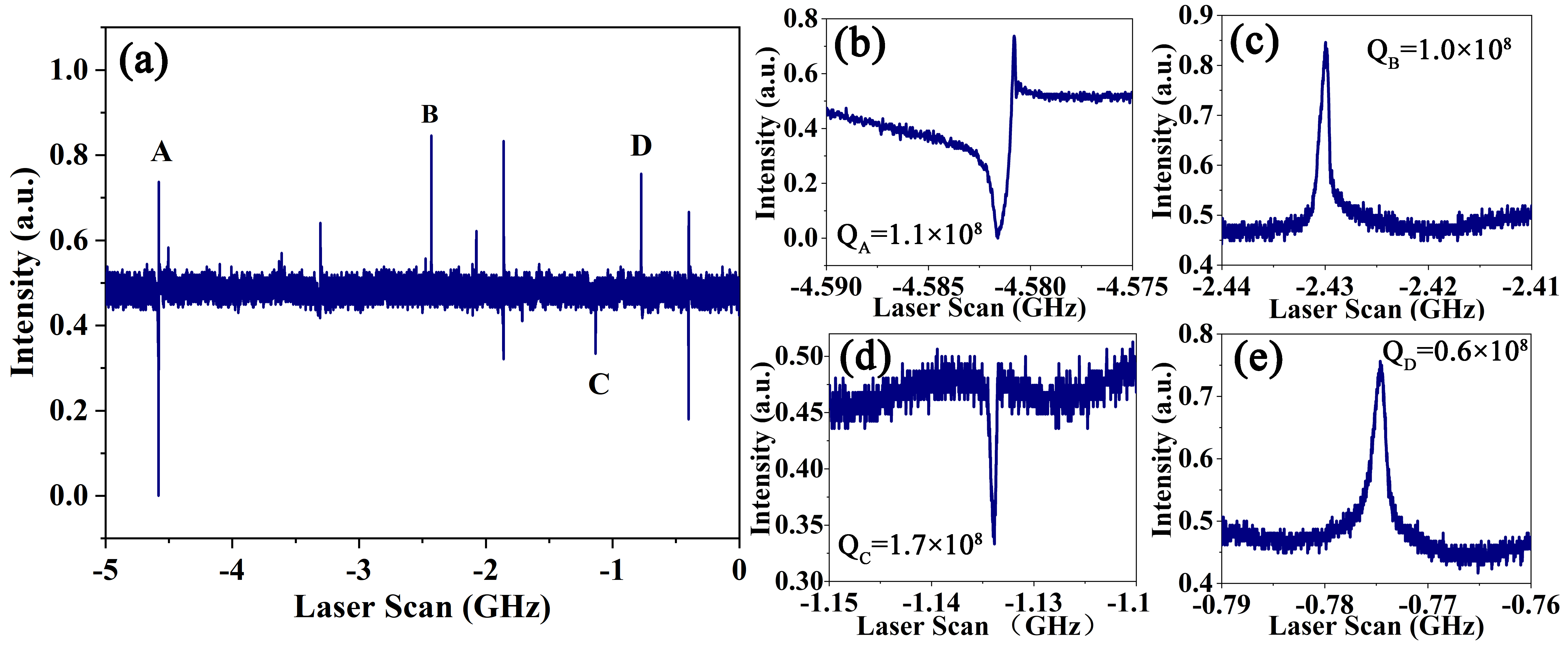}
\caption{WGM spectrum. (a) Reflection spectrum under a 1531 nm tunable laser scan. (b)-(e) Details of the four labeled resonances in the reflection spectrum. }
\label{fig:WGM}
\end{figure}

The pump source was a tunable laser  (TLB 6700, Newport) operating near 1531 nm. Through excitation via the nanoantenna, we observed a WGM spectrum containing multiple resonances (Fig. \ref{fig:WGM} (a)). Four typical resonances are labeled in the spectrum as A, B, C, and D, and their details are shown in Fig. \ref{fig:WGM}(b)-(e), respectively. We see that A is a typical Fano resonance (\begin{math}Q_{\rm A}=1.1\times10^8\end{math}) formed by a continuous spectrum corresponding to the field reflected from the surface of the sphere and a discrete spectrum corresponding to the scattering cavity field \cite{Shu:12}. The modes B, C, and D correspond to three typical Lorentz peaks or dips and their Q-factors are \begin{math}Q_{\rm B}=1.0\times10^8\end{math}, \begin{math}Q_{\rm C}=1.7\times10^8\end{math} and \begin{math}Q_{\rm D}=0.6\times10^8\end{math}. We see that the resonator maintains high Q-factors after packaging.

\begin{figure}[htbp]
\centering\includegraphics[width=11cm]{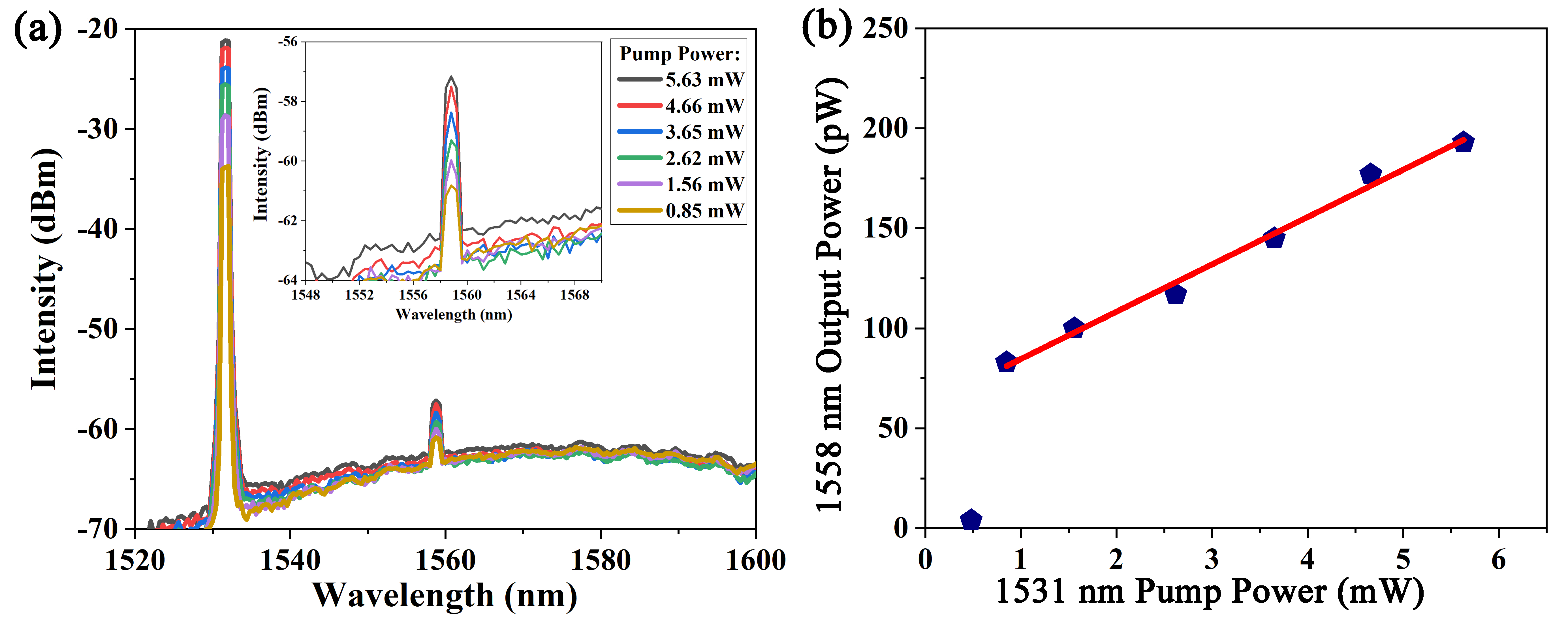}
\caption{Laser performance of an Er$^{3+}$-doped packaged microsphere device. (a) The laser spectrum at different 1531~nm pump powers. The lasing emission is centered on 1558~nm. (b) Output laser power at 1558~nm as a function of 1531~nm pump power light. Blue dots: experimental data; Red line: linear fit. The lasing threshold is at around 0.85~mW.}
\label{fig:laser}
\end{figure}

We have also prepared an active packaged device using the same method as described for the passive device but Er$^{3+}$-doped fiber (M12-980-125, Thorlabs) to make the microspheres. When 1531 nm pump light was coupled into the resonator via the nanoantenna, lasing behavior was observed at 1558 nm. Figure \ref{fig:laser}(a) shows the output laser spectrum at different pump powers. The relationship between the pump power and the lasing output power is shown in Fig. \ref{fig:laser}(b). Note that the pump power is measured at the input end of the nanoantenna fiber. The launched pump power at lasing threshold is less than 0.85 mW and the laser power shows a linear relationship with the pump power above threshold. The launched pump power is of the same order as that recorded  for lasing in a WGM resonator by pumping via a Rayleigh scatter, that is 0.485 mW \cite{2014Interfacing}. In our experiment, the laser displayed single-mode operation without any sign of saturation when the pump power was increased to 5.63 mW. At this input power the back reflected pump power measured at the input to the OSA with a power meter was 7.7 $\mu$W. Assuming the input and output coupling strengths are equal, this equates to a coupling efficiency of 11.4 $\%$. At threshold, the input power can then be expected to be 93 $\mu$W. Threshold could be lowered by locking the pump laser to a WGM or optimizing the pump polarization and tip position.

\section{Conclusion}

In conclusion, we have designed and fabricated a packaged WGM device based on a fiber nanoantenna as a coupler for both a passive and an active microsphere resonator. The device has the advantages of having a small footprint and good resistance to environmental interference and pollution whilst maintaining high Q-factors of the order of 10$^8$ for the passive devices.  Since the Rayleigh scattering coupling mechanism does not rely on phase matching \cite{1588878} this scheme could, in principle, be used to fabricate packaged WGM devices from a range of different materials for the resonator, such as multi-component glass, crystals, etc. Another important point to note is that the nanoantenna can achieve Rayleigh-induced coupling without the need to deposit a scatterer onto the surface of the WGM resonator which is a non-trivial task if one wishes to do so in a deterministic manner. The point-and-play feature of the nanoantenna  simplifies the coupling method significantly. In summary, this work offers a reliable way to provide environmental isolation, increased robustness, and portability to WGM resonators. This will pave the way to move WGM resonators from the laboratory to the field for a variety of applications such as low threshold lasers, filters, and sensors.

\section*{Funding}
National Key Program of the Natural Science Foundation of China (NSFC) (61935006, 62090062); China Scholarship Council (201906680065); The 111 project (B13015) to the Harbin Engineering University; Okinawa Institute of Science and Technology Graduate University; \textcolor{red}{Science and Engineering Research Board, Department of Science and Technology, India, under Grant  CRG/2019/001215.}

\section*{Acknowledgments}
The authors acknowledge the Engineering Support Section of OIST Graduate University.

\section*{Disclosures}
The authors declare no conflicts of interest.

%%%%%%%%%%%%%%%%%%%%%%% References %%%%%%%%%%%%%%%%%%%%%%%%%

\bibliography{Ref}

\end{document}